\documentclass[aps,twocolumn,showpacs,preprintnumbers,nofootinbib,prd,10pt,superscriptaddress,longbibliography,firstinits]{revtex4-2}

\usepackage{graphicx,amssymb,amsmath,amsthm,amsfonts,epsfig,epsf,fixmath}
\usepackage{comment}
\usepackage{mathrsfs}
\usepackage[usenames]{color}
\usepackage{epstopdf}
\usepackage{dcolumn}
\usepackage{float}
\usepackage{latexsym}
\usepackage{rotating}
\usepackage{longtable}

\usepackage{enumerate}
\usepackage{tensor,multirow}
\usepackage{url}
\usepackage{bm}
\usepackage[dvipsnames]{xcolor}
\usepackage[unicode]{hyperref}
\hypersetup{colorlinks=true, citecolor=MidnightBlue,
            linkcolor=Maroon, urlcolor=MidnightBlue, linktocpage=true}

\newcommand{\hM}{\hat{M}}
\newcommand{\hQ}{$\hat{Q} \ $}

\newcommand{\hMthr}{$\hat{M}_{thr} \ $}

\newcommand{\bbcrit}{$\gamma_{crit}$}
\newcommand{\bb}{$\gamma$}

\newcommand{\freg}{F_{\rm reg}}

\begin{document}

\title{\textbf{Minimum mass, maximum charge and hyperbolicity in scalar Gauss-Bonnet gravity}}

\author{Dario Rossi}
\email{dario.rossi@phd.unipi.it}
\affiliation{Dipartimento di Fisica, Università di Pisa, Largo B. Pontecorvo 3, 56127 Pisa, Italy}
\affiliation{INFN, Sezione di Pisa, Largo B. Pontecorvo 3, 56127 Pisa, Italy}
\author{Leonardo Gualtieri}
\email{leonardo.gualtieri@unipi.it}
\affiliation{Dipartimento di Fisica, Università di Pisa, Largo B. Pontecorvo 3, 56127 Pisa, Italy}
\affiliation{INFN, Sezione di Pisa, Largo B. Pontecorvo 3, 56127 Pisa, Italy}
\author{Thomas~P.~Sotiriou}
\email{thomas.sotiriou@nottingham.ac.uk}
\address{Nottingham Centre of Gravity,
Nottingham NG7 2RD, United Kingdom}
\address{School of Mathematical Sciences, University of Nottingham,
University Park, Nottingham NG7 2RD, United Kingdom}
\address{School of Physics and Astronomy, University of Nottingham,
University Park, Nottingham NG7 2RD, United Kingdom}

\begin{abstract}

    We study the loss of hyperbolicity of perturbation equations for  black hole solutions of scalar Gauss-Bonnet gravity. We consider a class of coupling functions allowing for static black hole solutions with arbitrary small masses. For masses below a minimum value, such solutions become unphysical, because the perturbation equations become elliptic; this arguably corresponds to the loss of validity of the effective field theory. We analyse  the dependence of this minimum mass on the parameters of the theory, finding that with an appropriate choice of the coupling function, such mass can be chosen arbitrarily small. However, this does not correspond to larger deviations from general relativity, since observable quantities like the black hole scalar charge are bounded by above.
\end{abstract}

\maketitle

\section{Introduction}
The detection of gravitational wave (GW) signals from compact binary inspirals has opened a new observational window on the strong-field regime of gravity. It is now possible to test whether general relativity (GR), which successfully describes gravity in the weak-field regime, also provides an equally faithful description of gravitational phenomena when the gravitational fields are strong and the curvature is large.

Scalar-tensor theories, in which one or more scalar fields are non-minimally coupled to gravity, are arguably the simplest GR modifications which can be tested by GW detectors. Among them, we shall consider scalar-Gauss Bonnet (sGB) gravity theories, in which a real scalar field $\phi$ is coupled to the Gauss-Bonnet (GB) invariant 
\begin{equation}
\label{eq:GB_term}
    \mathcal{G}= R^2-4 R_{\mu \nu} R^{\mu \nu} + R_{\mu \nu \alpha \gamma} R^{\mu \nu \alpha \gamma},
\end{equation}
where $R_{\mu \nu \alpha \gamma}$, $R_{\mu \nu}$ and $R$ are the Riemann tensor, the Ricci tensor and the Ricci scalar, respectively. The sGB action is (in geometric units $G=c=1$)
\begin{equation}
\label{eq:sGB_action}
        S = \frac{1}{2}\int d^4x \sqrt{-g}\biggl(R-\frac{1}{2}\nabla_\mu \phi \nabla^\mu \phi +V(\phi) + \alpha f(\phi) \mathcal{G} \biggr)~,
\end{equation}
where $V(\phi)$ is the scalar field potential (which we shall neglect for simplicity), and $\alpha$ is the coupling constant of the sGB interaction, with dimensions $({\rm lenght})^2$, providing the length-scale of the interaction. 

This class of theories is particularly interesting, for several reasons. First of all, they are the simplest scalar-tensor theories of gravity with second-order in time field equations (and thus without an Ostrogradsky ghost~\cite{berti2015testing}) which can evade the no-hair theorems~\cite{Kanti:1995vq,Sotiriou:2013qea,Sotiriou:2014pfa}, allowing for stationary BH solutions different from those of GR. Second, they naturally arise in fundamental physics frameworks, such as string theory~\cite{Gross:1986mw} or effective field theory~\cite{Weinberg:2008hq}. Finally, in recent years the waveform of a binary BH coalescence in sGB gravity has been computed through the  inspiral~\cite{Yagi:2011xp,Julie:2019sab,Shiralilou:2021mfl,Julie:2024fwy}, merger~\cite{East:2020hgw,East:2021bqk,AresteSalo:2022hua,Doneva:2023oww,Corman:2022xqg,Doneva:2024ntw,Corman:2025wun} and ringdown~\cite{Pierini:2022eim,Chung:2024vaf,Chung:2025gyg,Khoo:2024agm,Blazquez-Salcedo:2024dur} stages; presently, this is the only example of a full waveform in a theory beyond GR.

In recent years, different theories of this class -~corresponding to different choices of the coupling function $f(\phi)$~- have been studied, such as the exponential coupling ($f=e^\phi$) of Einstein-dilaton GB gravity~\cite{Kanti:1995vq}, or the linear coupling ($f=\phi$) of  shift-symmetric GB gravity~\cite{Sotiriou:2013qea, Sotiriou:2014pfa}. In these theories, the BHs are necessarily different from those of GR, and have non-trivial scalar field profiles (we call such solutions {\it hairy BHs}). Other sGB theories, in which $f(\phi)$ has a stationary point (e.g. the quadratic $f=\phi^2$~\cite{Silva:2017uqg} or the Gaussian $f=1-e^{-\frac{3}{2}\phi^2}$~\cite{Doneva:2017bvd} couplings), allow both Kerr BHs and hairy BHs; in this case the process of {\it spontaneous scalarization} can occur, in which a Kerr BH dynamically grows a scalar field profile~\cite{Doneva:2017bvd,Silva:2017uqg,Dima:2020yac,Herdeiro:2020wei,Doneva:2022ewd}.

A well-posed initial value formulation exists for sGB gravity theories~\cite{Kovacs:2020ywu}, under the assumption of weak coupling. For BH dynamics, this assumption is satisfied when the horizon radius $r_{\rm h}$ (i.e., in geometric units, the mass $M$) is sufficiently smaller than the characteristic length-scale associated with $\alpha$. Indeed, non-linear simulations of dynamical BH processes have demonstrated hyperbolicity loss for larger values of the coupling, with elliptic regions appearing during the evolution~\cite{Ripley:2019hxt,East:2020hgw,AresteSalo:2022hua}. It has been suggested that this loss of hyperbolicity is related to the fact that sGB gravity is only consistent as the low energy limit of an effective field theory expansion, and loses validity for processes with energy scales larger than the characteristic scale of the expansion~\cite{Serra:2022pzl,R:2022hlf}. It has also  been suggested that additional interactions could help mitigate well-posedness issues~\cite{Thaalba:2023fmq,Thaalba:2024htc,Thaalba:2024crk}.

A remarkable feature of BHs in sGB gravity is the existence of a minimum mass. For theories with $f'\neq0$ (i.e., not allowing spontaneous scalarization), a minimum mass of stationary BHs  appears to be a generic feature~\cite{Kanti:1995vq,Kleihaus:2011tg,Sotiriou:2014pfa,Thaalba:2022bnt,Thaalba:2025lwe}. For theories that allow for spontaneous scalarization, stationary BH solutions 
can exist for all values of the mass for certain types of couplings (e.g.~\cite{Doneva:2017bvd}); however, when the mass is smaller than a threshold value, the equations for radial perturbation cease to be hyperbolic~\cite{Blazquez-Salcedo:2018jnn}. In such cases, a minimum mass for BHs arises from hyperbolicity requirements.

The main aim of this article is to explore the bounds on the minimim mass set by hyperbolicity. We shall focus on a particular class of coupling functions, the so-called {\it general Gaussian couplings} introduced in~\cite{Fernandes:2022kvg} (see also~\cite{Doneva:2021tvn}). In particular, we shall show that the threshold mass associated to hyperbolicity can be arbitrarily small for these theories.
We shall also study the impact that an additional coupling to the Ricci scalar can have on hyperbolicity-induced bounds on the minimum mass. Scalar GB gravity with Ricci coupling (sGBR gravity) has been shown to render GR a cosmological attractor~\cite{Antoniou:2020nax} and  improve both stability~\cite{Antoniou:2021zoy,Antoniou:2022agj} and hyperbolocity~\cite{Antoniou:2022agj,Thaalba:2023fmq}.

Part of the motivation for this analysis is of theoretical nature --- understand the link between the minimum mass, hyperbolicity and the range of validity of sGB gravity as an EFT --- and part is related to observations. One can express observational bounds in terms of the dimensionless ratio (see e.g.~\cite{Pani:2009wy,Maselli:2015tta,Blazquez-Salcedo:2016enn}) 
\begin{equation}
    \label{defzeta}\zeta=\frac{\alpha}{M^2}~.
\end{equation}
A lower bound on $M$ corresponds to an upper bound on $\zeta$ (for a given coupling $\alpha$), and then to a theoretical upper bound to observable deviations.

A naive analysis of Eq.~\eqref{defzeta} suggests that a sGB theory with a smaller minimum mass would also allow to larger values of $\zeta$ and then to larger observable deviations from GR.
We will show that this is not always the case. For theories with general Gaussian couplings, although the upper bound on $\zeta$ can be arbitrarily large, this does not lead to large observable effects.

The structure of the paper is the following. 
In Sec.~\ref{sec:theo} we discuss static, spherically symmetric BHs in sGB gravity -~focussing on general Gaussian coupling functions~- and introduce the equations for radial perturbations. In Sec.~\ref{sec:results} we present our results concerning the minimum mass, the scalar charge and the coupling with the Ricci scalar. Finally, in Sec.~\ref{sec:concl} we discuss our results and draw our conclusions.


\section{Theoretical framework}\label{sec:theo}
\subsection{Action and field equations}\label{sec:Action}

We will consider the following action for sGBR gravity:
\begin{equation}
\label{eq:sGBR_action}
    S = \frac{1}{2}\int d^4x \sqrt{-g}\biggl(R-\frac{1}{2}\nabla_\mu \phi \nabla^\mu \phi +J(\phi)R + \alpha f(\phi) \mathcal{G} \biggr)~,
\end{equation}
where the dimensionless function $J(\phi)$ describes the coupling of the scalar field with the Ricci scalar. The corresponding field equations are~\cite{Antoniou:2022agj}
\begin{subequations}
\label{eq:sGBR_field_equations}
\begin{align}
    \label{eq:sGBR_fields_equations_1}
    \left(J(\phi)+1\right)G_{\mu\nu}&= T^\phi_{\mu\nu} - \mathcal{K}_{\mu\nu} + \\ \nonumber & \quad + \left(\nabla_\mu \nabla_\nu - g_{\mu\nu}\Box\right)J(\phi)\\
    \label{eq:sGBR_fields_equations_2}
    \Box \phi ~&= -  \alpha f'(\phi) \mathcal{G}-J'(\phi)R\,, 
\end{align}
\end{subequations}
where
\begin{align}
    \label{eq:T_munu}
   T^{\phi}_{\mu\nu}&=\frac{1}{2}\partial_\mu \phi \partial_\nu \phi - \frac{1}{4}g_{\mu\nu}(\partial_\alpha \phi)^2\\
   \mathcal{K}_{\mu\nu}&= \alpha g_{\alpha(\mu}g_{\nu)\gamma}~
   \epsilon^{\rho\gamma\sigma\beta}\nabla_{\lambda}\bigl[~^{*}\tensor{R}{^\alpha^\lambda_\sigma_\beta}f'(\phi)\nabla_\rho \phi\bigr].
   \label{eq:K_munu}
\end{align}
The Kerr metric is solution as long as $f'(\phi_0)=J'(\phi_0)=0$ 
for some constant $\phi_0$. The equation for sGB gravity are recovered for $J\equiv0$.

\subsection{Background solutions}\label{subsec:background}

As mentioned in the Introduction, we can define two classes of sGB (or sGBR) gravity theories: those for which the condition
\begin{equation}
 \label{eq:GB_conditionToGR}
     f'(\phi_0)=0
 \end{equation}
is not satisfied for any $\phi_0$, and those for which it can be satisfied.

For theories with $f'\neq0$~\cite{Kanti:1995vq,Kleihaus:2011tg,Sotiriou:2014pfa}, stationary BH solutions always have a non-trivial scalar field profile. The field equations have one stationary solution for each choice of mass $M$ and angular momentum $Ma$, provided the mass is larger than a minimum value $M_{\rm min}$, which is proportional to $\sqrt{\alpha}$ (i.e., that $\zeta<\zeta_{\rm max}=\alpha/M^2_{\rm min}$), and of course that the angular momentum satisfies the (generalized) Kerr bound. For instance, in the case of Einstein-dilaton GB gravity, $\zeta_{\rm max}\simeq0.619$ for non-rotating BHs~\cite{Pani:2009wy}; for rotating BHs, and for different choices of the coupling functions in this class, the value of $\zeta_{\rm max}$ is different, but always $\sim O(1)$. As shown in~\cite{Sotiriou:2014pfa}, the singularity of BHs in sGB gravity theories is located at a constant radius surface, which reaches the horizon as $M\to M_{\min}$, therefore a naked singularity appears for smaller masses.

The situation is different for the theories satisfying the condition~\eqref{eq:GB_conditionToGR}, which will be the focus of this paper. In this case, for each value of $M$ and of $a$ ($|a|\le M$) the Kerr metric is solution of the field equations~\eqref{eq:sGBR_field_equations}, with constant scalar field $\phi=\phi_0$.
 In part of the parameter space $(M,a)$, additional (hairy) solutions exist~\cite{Doneva:2017bvd,Antoniou:2017acq,Herdeiro:2020wei,Berti:2020kgk}. In these regions of the parameter space, (stable) hairy BH solution can be dynamically connected to the Kerr solution via a tachyonic instability associated with the  process of spontaneous scalarization, see Ref.~\cite{Doneva:2022ewd} for a review.
 
We remark that in scalarization models stationary BHs do not have a minimum mass, as  Kerr BHs are  solutions for any mass. In certain models, stationary hairy solutions can have a minimum mass (e.g.~\cite{Silva:2017uqg, Antoniou:2021zoy}) while in other models they can have arbitrarily low mass (e.g.~\cite{Doneva:2017bvd}).  However, as we shall discuss in Sec.~\ref{subsec:pert}, BHs in these models can become unstable,  and the equations governing the dynamics of their perturbations can cease to be hyperbolic, when the BH mass lies below a certain threshold $M_{\rm thr}$. Hence, a minimum mass bound can arise by considering the dynamics of perturbed BHs.

\subsubsection*{Static spherically symmetric background}

In this paper we shall not consider the case of a rotating background, but instead focus on static, spherically symmetric BH  spacetimes, which can take the form
\begin{equation}
    \label{def:spherically_symmetric_metric}
    dS^2 = -A(r)dt^2 +  B(r)^{-1}dr^2 + r^2 (d\theta^2+\sin^2\theta d\varphi^2)~.
\end{equation}
With this ansatz, equations~\eqref{eq:sGBR_field_equations} give a system of coupled ordinary differential equations for $A(r)$, $B(r)$ and $\phi(r)$, which can be solved imposing regularity at the horizon and asymptotic flatness with $\phi\to\phi_0$ at infinity. The solution is characterized by the ADM mass $M$ and by the {\it scalar charge} $Q$, which can determined by the asymptotic ($r\to\infty$) solutions, as 
$B(r)\to 1-2M/r$, $\phi \to \phi_0+Q/r$.

As mentioned above, we shall consider coupling functions satisfying the condition 
$f'(\phi_0)=J'(\phi_0)=0$. The simplest choice is $f(\phi)\propto\phi^2$ with $J(\phi)=0$~\cite{Silva:2017uqg}, for which $\phi_0=0$. In this case while the Schwarzchild solution is allowed for any value of the mass, the hairy solution exists in some intervals of $M$, $M\in[M_0,M_1]\cup[M_2,M_3]\cup\cdots$, which {\it do not include} the limit $M\to0$. 
Moreover, for this choice of the coupling function, the scalarized branches are all unstable for radial perturbations~\cite{Blazquez-Salcedo:2018jnn}, which will be discussed in Sec.~\ref{subsec:pert}; the Schwarzschild solution, instead, is unstable for $M<M_0$. The stability of the solution can be improved by adding a self-interaction for the scalar field~\cite{Macedo:2019sem}, or a quartic term to the quadratic coupling~\cite{Silva:2017uqg}. However, even with such modifications, stationary solutions -~regardless of their stability~- are not allowed in the limit of small masses.


A different kind of coupling (in the same class satisfying Eq.~\eqref{eq:GB_conditionToGR}) is the Gaussian coupling~\cite{Doneva:2017bvd} $f\propto 1-e^{-\frac{3}{2}\phi^2}$.
In this case --~at least when $J(\phi)=0$~-- hairy BHs exist in an interval $M\in[0,M_1]$; therefore, both hairy and GR BHs exist for arbitrary small values of the mass. 
However, as mentioned above, only those with masses larger than a threshold value $M_{\rm thr}$, are radially stable. Summarizing, for $M>M_1$ only the Schwarzschild solution exsists, and is stable; for $M_{\rm thr}<M<M_1$ the Schwarzschild solution is unstable while the hairy solution is stable; for $0<M<M_{\rm thr}$ both the GR and hairy solutions are unstable.

We shall consider a generalization of the coupling of~\cite{Doneva:2017bvd}, the general Gaussian coupling functions~\cite{Doneva:2021tvn,Fernandes:2022kvg}:
\begin{equation}
\label{eq:exponentialcoupling}
    f(\phi) = \frac{1}{8 \gamma}\left(1-e^{- \gamma \phi^2}\right)~,
\end{equation}
which depend on the dimensionless parameter $\gamma$. As shown in~\cite{Fernandes:2022kvg} (see also Appendix~\ref{appendix:beta_crit}), when $\gamma$ is larger than a critical value $\gamma_{\rm crit}\simeq1.18$, the structure of the solution is the same discussed above for the  Gaussian coupling studied in~\cite{Doneva:2017bvd} (which is retrieved for $\gamma=3/2$~\footnote{Since our normalizations are different than those of other papers, the scalar field of this article differs to that of~\cite{Doneva:2017bvd} by an overall factor $2$, and to that of~\cite{Fernandes:2022kvg} by an overall factor $\sqrt{2}$. This also affects the normalization of $\gamma$.}): static, hairy BHs exist for arbitrary small masses. In the following we shall extend this analysis, studying the radial stability of these solutions.

Moreover, we shall generalize this analysis to sGBR gravity~\eqref{eq:sGBR_action}. Following Ref.~\cite{Antoniou:2022agj}, we choose a quadratic coupling  with the Ricci scalar:
\begin{equation}
\label{def:coupling_Ricci}
    J(\phi)=-\frac{\beta}{4}\phi^2~,
\end{equation}
where $\beta$ is a dimensionless coupling constant. 
A different choice (with a Gaussian functional form) will be discussed in Appendix~\ref{appendix:ricciexpcop}.
\subsection{Radial perturbations\label{subsec:pert}}

The scalar field and spacetime metric of a spherically symmetric BH undergoing radial perturbations have the general form~\cite{Silva:2018qhn,Blazquez-Salcedo:2018jnn}
\begin{align}
    \label{def:phi_perturbation}
    \phi &=\varphi_0(r) + \varepsilon \ \frac{\varphi_1 (t,r)}{r}\\
    \label{def:metric_func_perturbation}
    \begin{split}
        ds^2 &= - [A(r) + \varepsilon ~ F_t (t,r)]dt^2 + \\ &[B(r)^{-1} +\varepsilon ~ F_r (t,r)]dr^2 + r^2 d\Omega^2
    \end{split}
\end{align}
where $A \text{,} \ B \ \text{and} \ \varphi_0$ describe the background solution; $F_t $, $F_r$ and $\varphi_1$ are the first order perturbations of the $tt$ and the $rr$ components of the metric tensor and of the scalar field, respectively; $\varepsilon$ is a bookkeeping parameter for the perturbed analysis.
Expanding the unperturbed field equations~\eqref{eq:sGBR_field_equations} up to first order in $\varepsilon$ with the ansatz given in Eqs.~\eqref{def:phi_perturbation}, ~\eqref{def:metric_func_perturbation}, we obtain a system of five coupled partial differential equations (PDEs), which can be combined into a single second-order PDE for the radial perturbation of the scalar field, $\varphi_1(t,r)$:
\begin{equation}
\label{eq:perturbed_equation}
    \begin{split}
    g^2(r) & \frac{\partial^2 \varphi_1(t,r)}{\partial t^2} - \frac{\partial^2 \varphi_1(t,r)}{\partial r^2} + \\ &+ k(r)\frac{\partial \varphi_1(t,r)}{\partial r} + u(r) \varphi_1(t,r)=0~,
    \end{split}
\end{equation}
where $u(r), g(r)$ and $k(r)$ depend on the background metric  and scalar field. Full expressions of the coefficients can be found in~\cite{Blazquez-Salcedo:2018jnn} for sGB gravity, while for sGBR gravity they are shown in the \textsc{Mathematica} notebook in the Supplemental Material.
The stability of the background can then be studied by looking for exponentially growing modes, see e.g.~\cite{Blazquez-Salcedo:2018jnn,Macedo:2019sem,Antoniou:2022agj}. As noted in~\cite{Blazquez-Salcedo:2018jnn}, this instability can be considered as a consequence of the fact that, for certain values of the BH mass which depend on the coupling function, the function $g^2(r)$ appearing in the perturbation equation~\eqref{eq:perturbed_equation} becomes negative in a localized region outside the event horizon $r=r_{\rm h}$. As a consequence, the perturbation equation takes an elliptic character in that region, making the Cauchy problem ill-posed.

The requirement that the theory remains predictive as an EFT, can be used to impose a lower bound on the BH mass, as discussed in the Introduction. BHs whose perturbations do not satisfy hyperbolic equations cannot be the endpoint of a dynamical process and hence are not expected to exist.  

\section{Results}\label{sec:results}
\subsection{Hyperbolicity of BH perturbations}
We will now study the hyperbolicity of Eq.~\eqref{eq:perturbed_equation}, for sGB gravity and sGBR gravity with the general Gaussian coupling function~\eqref{eq:exponentialcoupling}. To this aim, we have verified whether  the function $g^2(r)$ is positive definite, for different values  of the dimensionless constants $\gamma$, $\beta$, and for different values of the mass of the static BH background.
We have considered $\gamma>\gamma_{\rm crit}$, for which (see Sec.~\ref{subsec:background}) hairy background solutions exist for arbitrary small values of the mass. An extended analysis of $\gamma_{\rm crit}$ is provided in Appendix~\ref{appendix:beta_crit}.

We find that the perturbation equations become elliptic for masses smaller than a threshold value $M_{\rm thr}$. In the following we discuss how this value depends on $\gamma$ and $\beta$ in sGB gravity and in sGBR gravity.

\begin{figure}[t]
    \centering
        \includegraphics[width=0.8\linewidth]{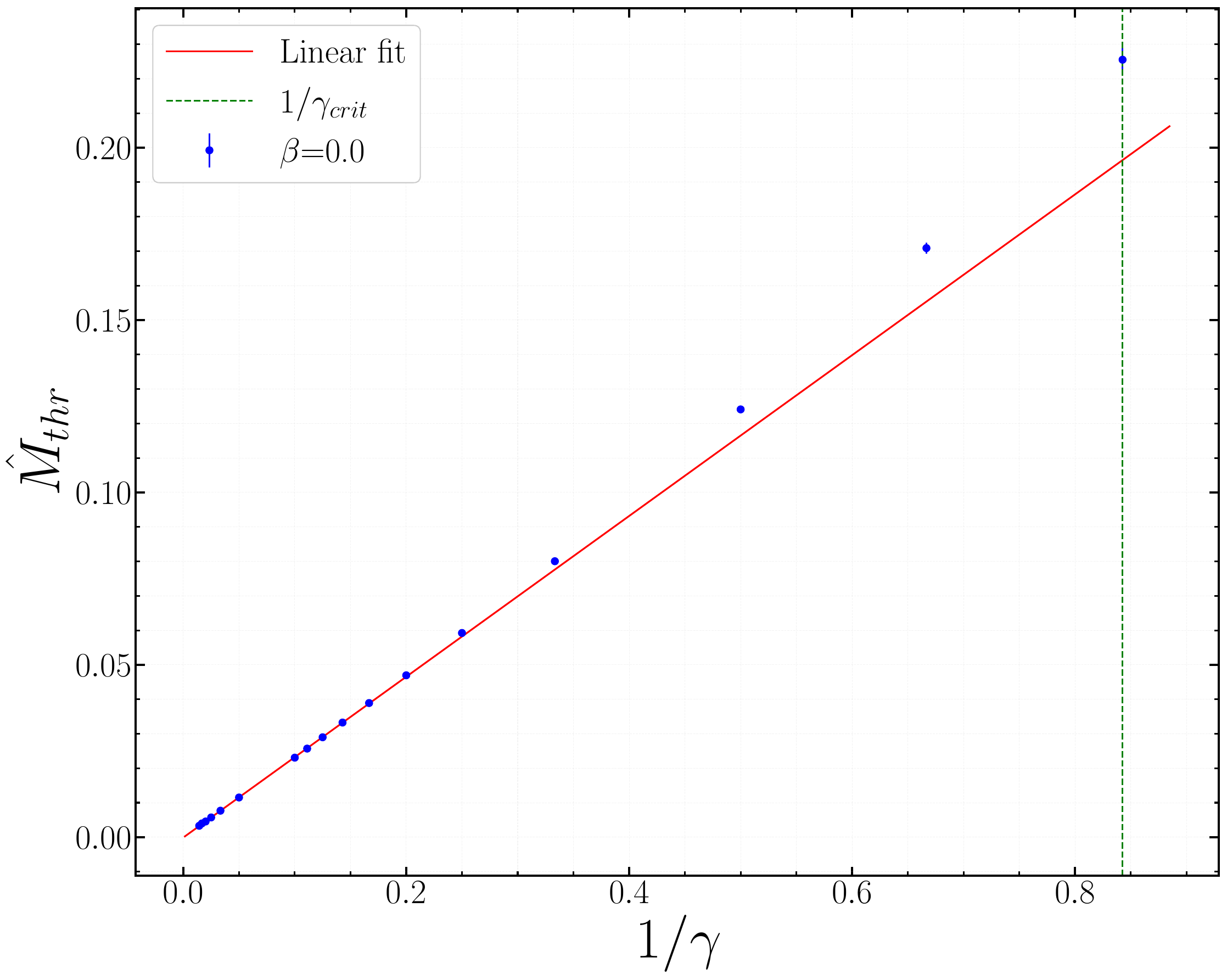}
    \caption{Threshold mass for hyperbolic equations, normalized with the coupling constant, as a function of $1/\gamma$. The red, solid line corresponds to the linear fit. The vertical green dashed line represents the critical value $\gamma_{\rm crit}$.}
    \label{fig:fit_ellipticity}
\end{figure}
\begin{figure}[t]
    \centering
    \includegraphics[width=0.8\linewidth]{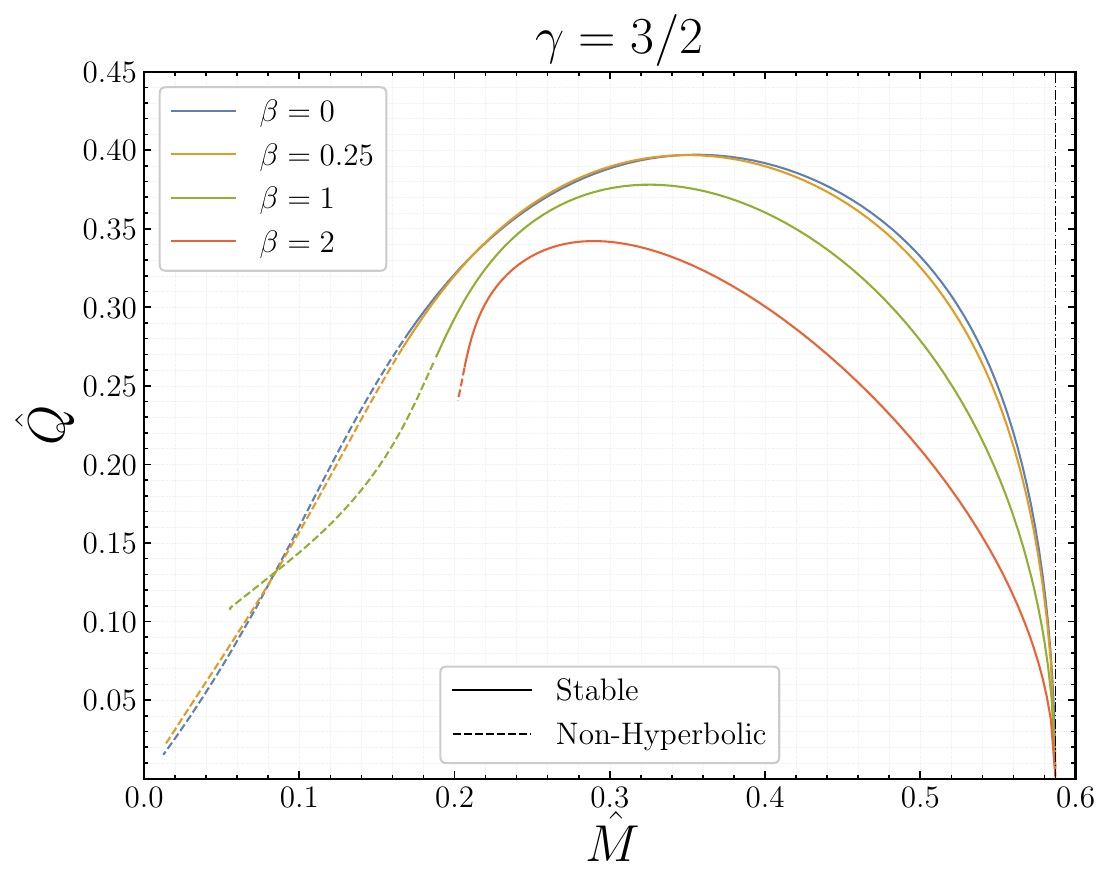}
    \includegraphics[width=0.8\linewidth]{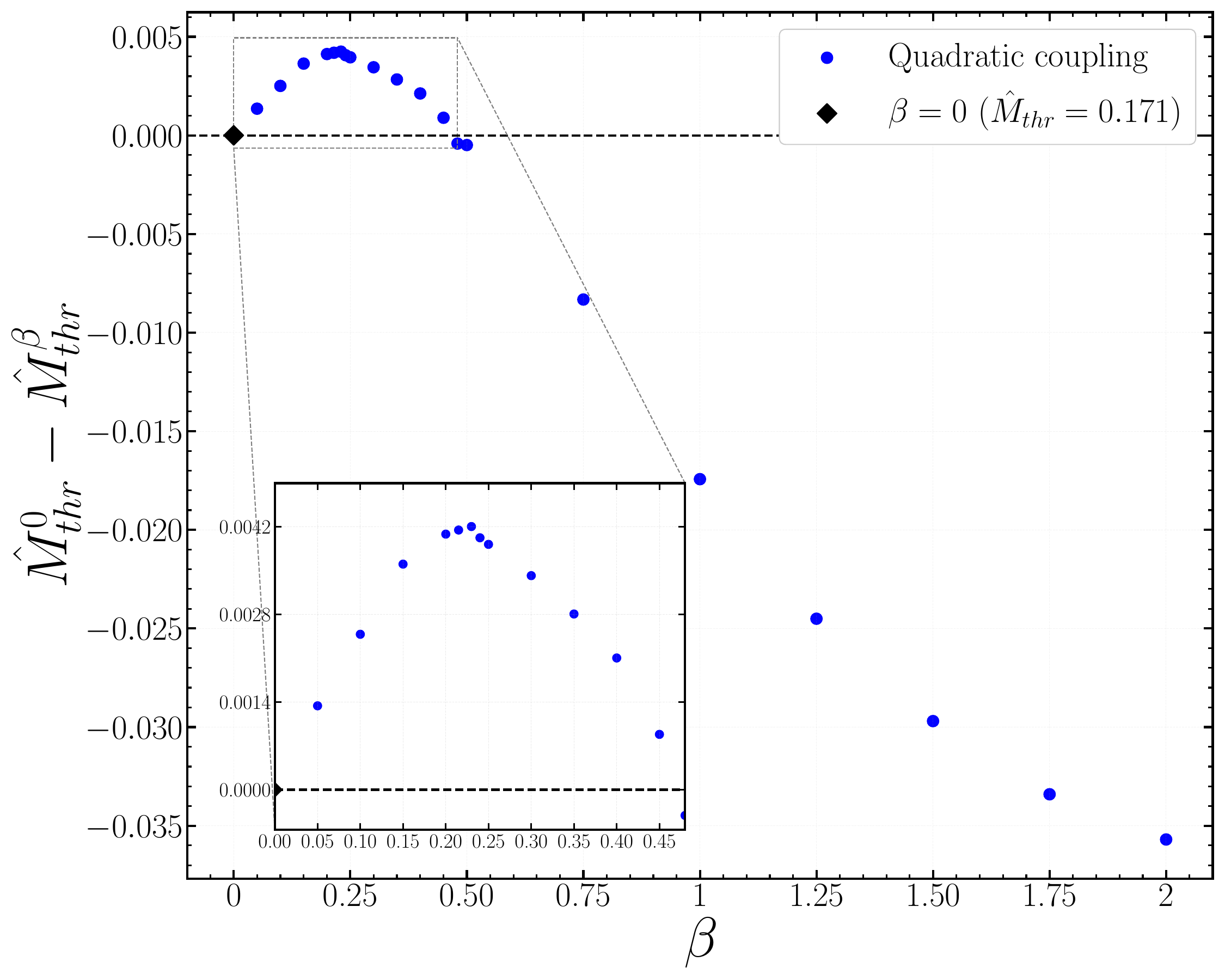}
    \caption{Top panel: dimensionless scalar charge \hQ~as a function of $\hM$~for fixed $\gamma=3/2$, with different values of $\beta$. Solid (dashed) lines represent hyperbolic (elliptic) configurations. The vertical dot-dashed line refers to maximum mass $M_1$ allowing for hairy solutions. 
    Bottom panel: Difference between the threshold mass in sGB and in sGBR gravity as a function of $\beta$, for $\gamma=3/2$. We can see that the Ricci coupling increases (decreases) the threshold mass for $\beta$ larger (smaller) than $\simeq0.50$.}
    \label{fig:QvsM_stability}
\end{figure}
\begin{figure}[t]
    \centering
    \includegraphics[width=0.8\linewidth]{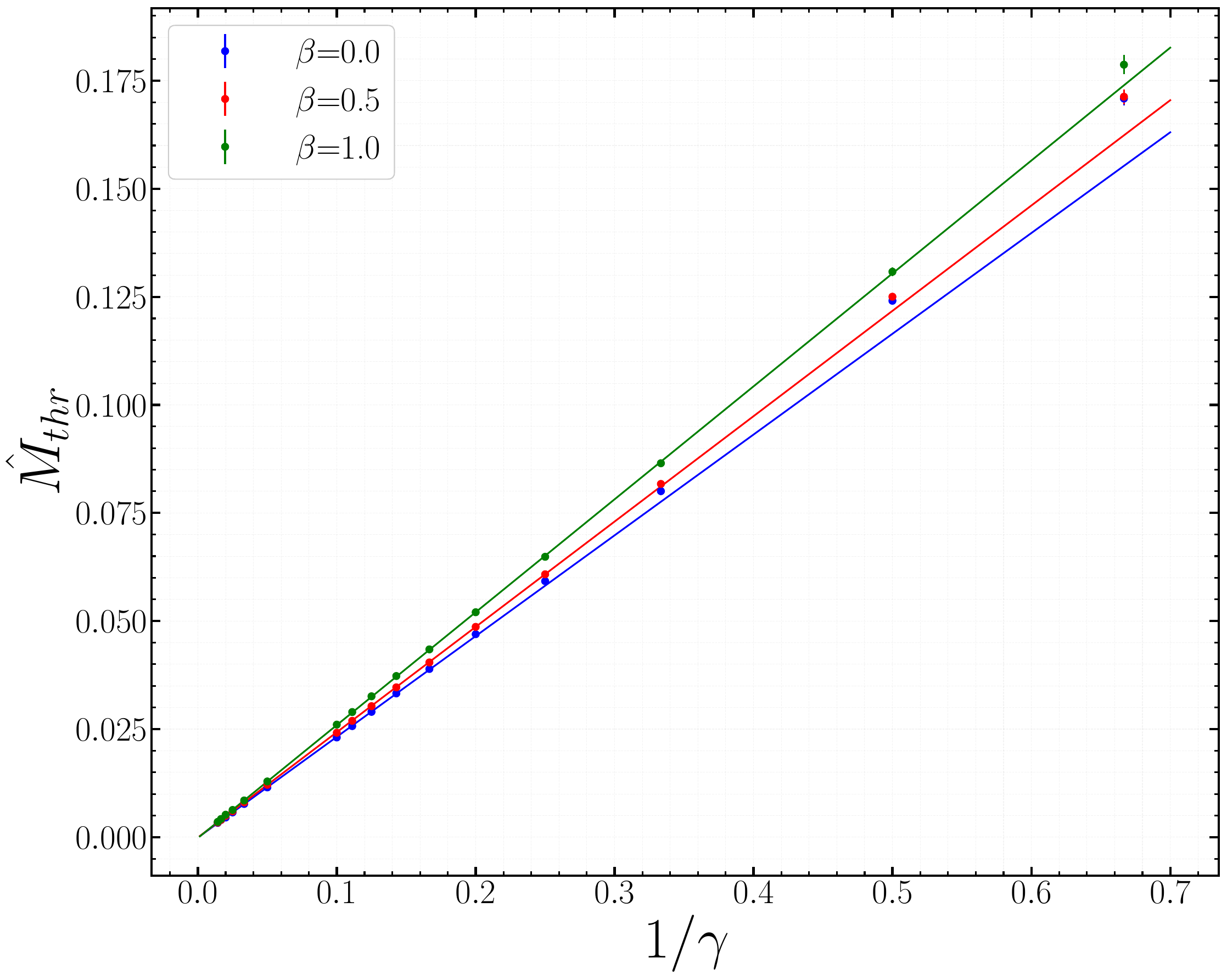}
    \caption{As in Fig.~\ref{fig:fit_ellipticity}, for different values of the $\beta$.}
    \label{fig:multiple_fits}
\end{figure}

\subsubsection*{sGB gravity}
In the case of sGB gravity ($\beta=0)$ we find that the threshold mass is a decreasing function of $\gamma$. In Fig.~\ref{fig:fit_ellipticity} we show the threshold mass normalized with the dimensionful couping constant, $\hat M_{\rm thr}=M_{\rm thr}/\sqrt{\alpha}$, as a function of $1/\gamma$. We can see that this behaviour is with good approximation linear. A linear fit of this curve gives $\hat M^{\rm fit}_{\rm thr}(\gamma)= b \gamma^{-1} + c$, with $b = 0.2337$ and $c = -0.00012$. Thus,
\begin{equation}
    \label{eq:fit_equation}
  \hat M_{\rm thr}\simeq\frac{b}{\gamma}~.  
\end{equation}
This suggest that it is possible to extend the regime of well-posedness of the theory to arbitrary small BH masses by considering a sufficiently large \bb. Note that for $\gamma\to\infty$, the theory reduces to GR.

\subsubsection*{sGBR gravity}

We now extend our hyperbolicity analysis to the case in which a Ricci coupling is present.

To begin with, we consider a given value $\gamma=3/2$ (corresponding to the case studied in~\cite{Antoniou:2022agj}). In Fig.~\ref{fig:QvsM_stability}  we show, in the top panel, the dimensionless scalar charge $\hat{Q}=Q/\sqrt{\alpha}$ as a function of $\hM$, for different values of $\beta$. The solid (dashed) curves correspond to hyperbolic (elliptic) configurations. This plot is consistent with the results of~\cite{Antoniou:2022agj}. 
%
We can see that  in most cases the threshold mass \hMthr  increases with $\beta$ (as noted in~\cite{Antoniou:2022agj}). However, for small values of $\beta$, the Ricci coupling has the opposite effect. This is clarified in the bottom panel, in which we show the shift in the threshold mass due to the Ricci coupling, as a function of $\beta$.

Finally, in Fig.~\ref{fig:multiple_fits} we show $\hat M_{\rm thr}$, as a function of $1/\gamma$, for different values of $\beta$. We can see that the linear behaviour also holds, with good approximation  in sGBR gravity, with the constant $b$ in Eq.~\eqref{eq:fit_equation} increasing with $\beta$. The linear fit becomes less accurate as $\gamma$ decreases.

The parameters of the fit  
\begin{equation}
 \hat M^{\rm fit}_{\rm thr}(\gamma,\beta)= b(\beta) \gamma^{-1} + c(\beta)\label{fitfunction}
\end{equation}
as functions of $\beta$ are shown in Table.~\ref{tab:fit_params}.
\begin{table}[t]
    \centering
    \begin{tabular}{c|c|c}
        \hline
        $\beta$ & $b $ & $c$ \\
        \hline
        0.0 & 0.2337 & -0.00012 \\
        0.25 & 0.2425 & -0.00021 \\
        0.4 & 0.2471 & -0.00023 \\
        0.5 & 0.260 & -0.0013 \\
        0.75 & 0.2669 & -0.00052 \\
        1.0 & 0.2833 & -0.00076 \\
        \hline
    \end{tabular}
    \caption{Best fit parameters of \hMthr as a function of $1/\gamma$ (Eq.~\eqref{fitfunction}) for different values of $\beta$.}
    \label{tab:fit_params}
\end{table}
\subsection{Charge per unit mass}
\begin{figure}[t]
    \centering
    \includegraphics[width=0.8\linewidth]{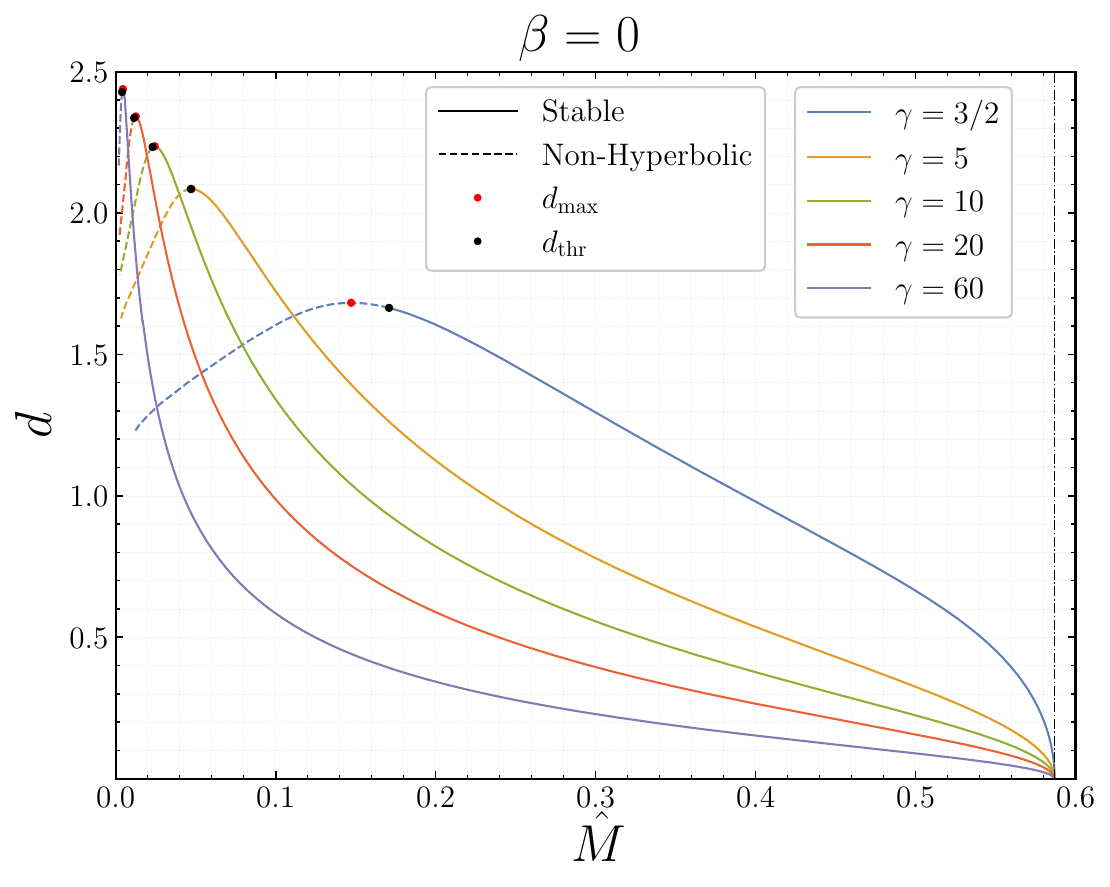}
    \caption{Dimensionless charge $d$ as a function of $\hM$, in sGB gravity with Gaussian coupling function and different values of $\gamma$. Black points ($d_{\rm thr}$) refer to the values of $d$ computed at the threshold masses \hMthr, separating stable branches (solid lines) from the non-hyperbolic regime (dashed lines). Red points ($d_{\rm max}$) correspond to the maximum value of $d$ for each $\gamma$.}
    \label{fig:QMvsM_alpha0}
\end{figure}
\begin{figure*}[t]
        \centering
    \includegraphics[width=0.405\textwidth]{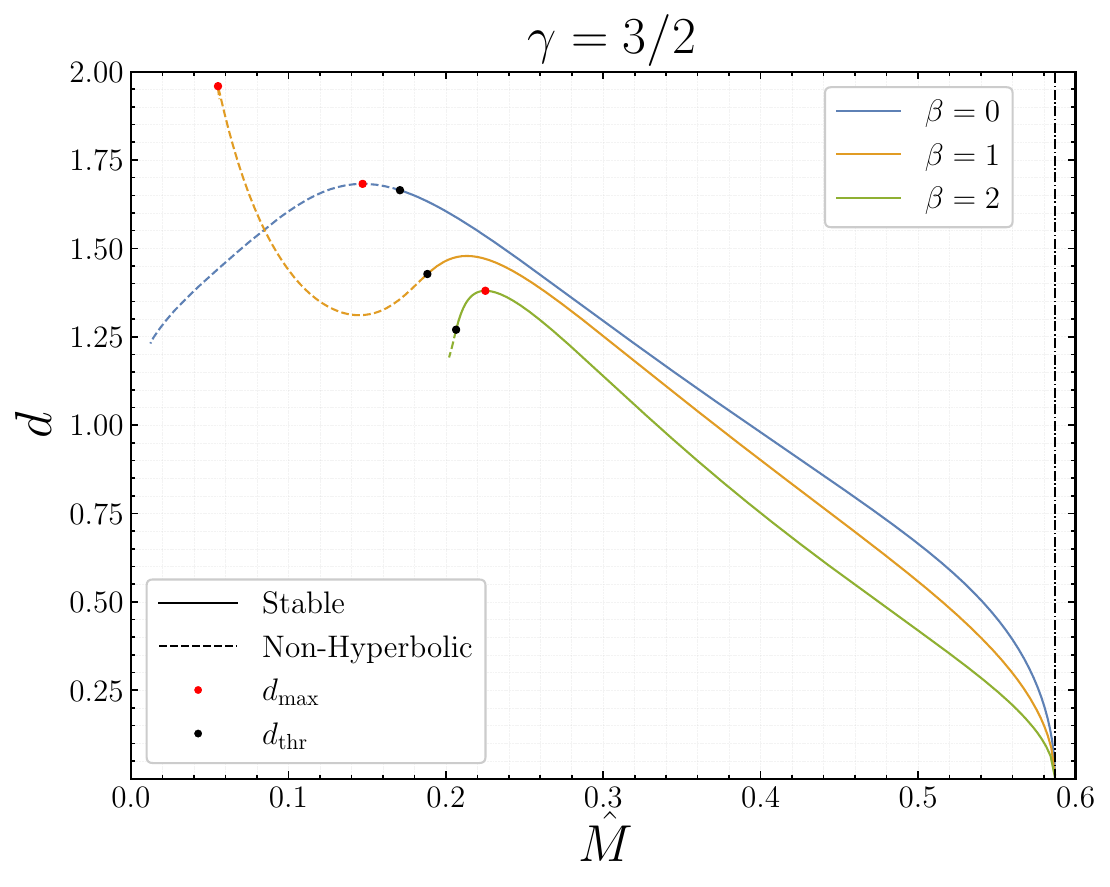}
    \includegraphics[width=0.405\textwidth]{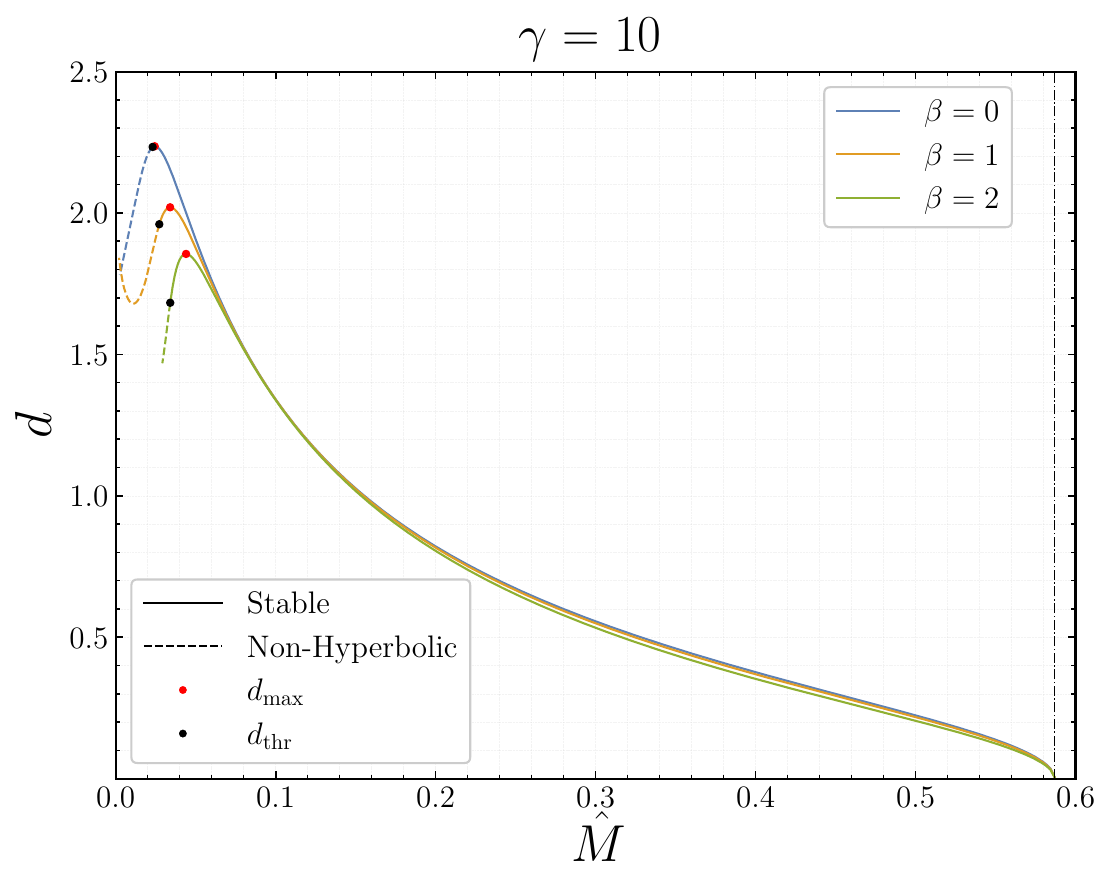}
    \caption{As in Fig.~\ref{fig:QMvsM_alpha0}, in sGBR gravity, for differnt values of $\beta$, for $\gamma=3/2$ (left panel) and $\gamma=10$ (right panel). }
    \label{fig:QMvsM}
\end{figure*}
\begin{figure*}[t]
        \centering
    \includegraphics[width=0.4\textwidth]{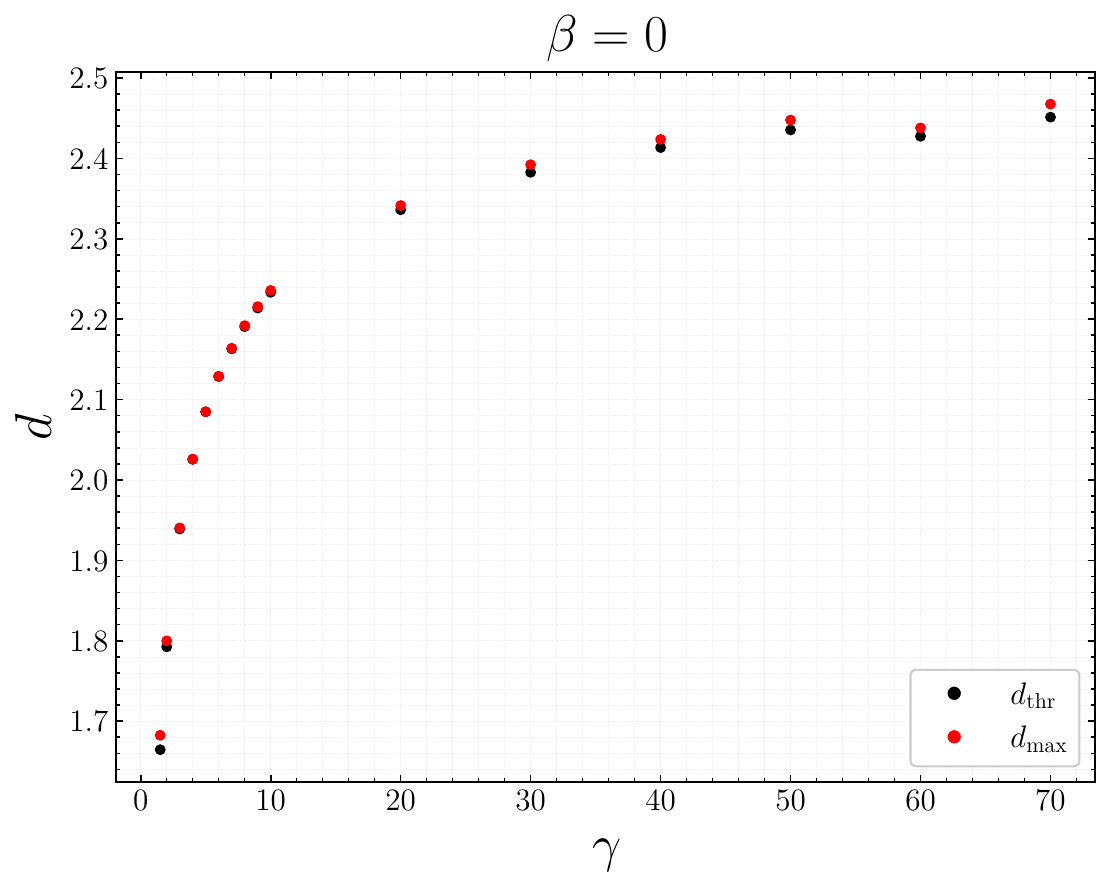}
    \includegraphics[width=0.4\textwidth]{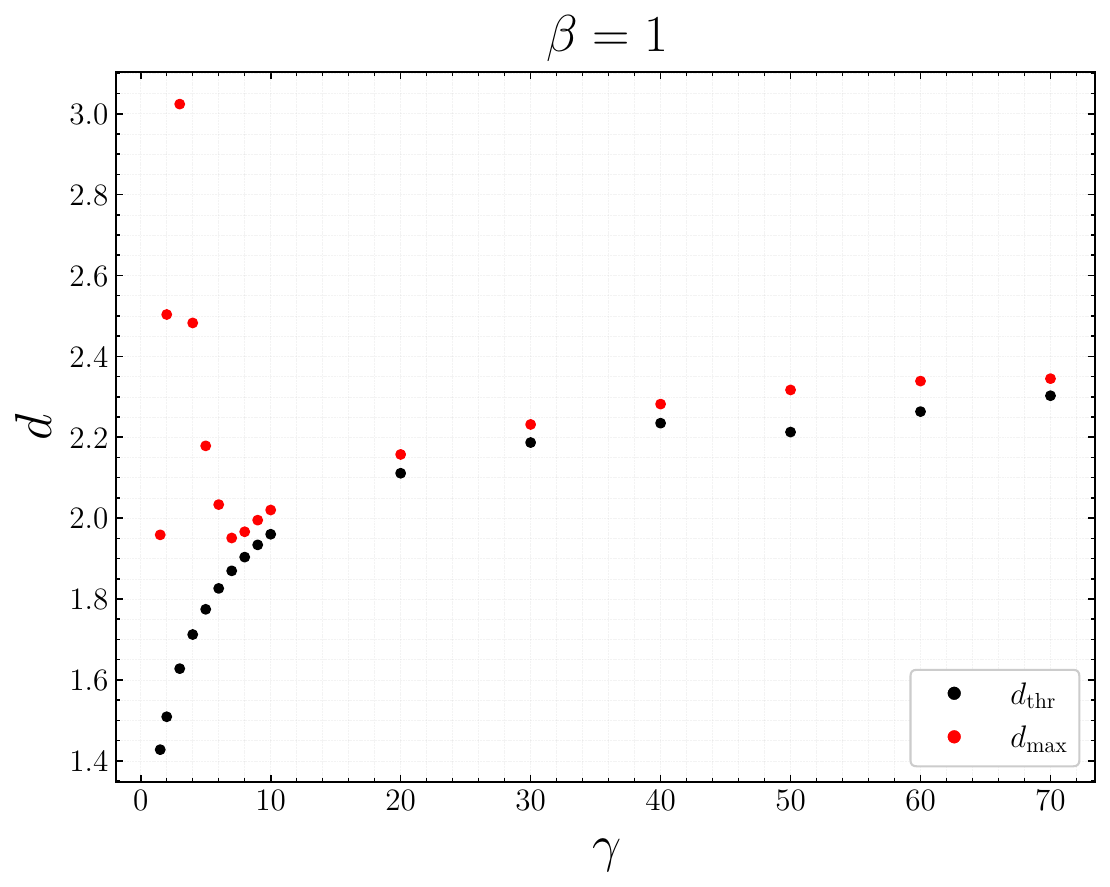}
    \caption{Maximum value of the dimensionless charge $d_{\rm max}$, and dimensionless charge at the threshold mass $d_{\rm thr}$, as functions of $\gamma$, for $\beta=0$ (left panel) and $\beta=1$ (right panel). }
    \label{fig:QMvsGamma}
\end{figure*}
Currently, the strongest constraints on sGB gravity come from the GW signal GW230529, observed by LVK in 2023, and emitted in the coalescence of a neutron star with a very light BH, of mass $M\simeq3.6\,M_\odot$. The analysis of this signal is consistent with GR, leading to an upper bound on the  fundamental length of the Gauss-Bonnet coupling, of the order of hundreds of meters. In particular, for sGB gravity with $f(\phi)=\phi$, the bound is $\sqrt{\alpha}\le 0.28$ km~\cite{Sanger:2024axs}. In this measurement, the analysis of the signal leads to a bound on the deviation of the $-1$PN component of the waveform, $\delta\hat\varphi_{-2}$, which is related to the difference of the {\it dimensionless scalar charge}
\begin{equation}
    d=\frac{Q}{M}\label{eq:dsc}
\end{equation}
of the two bodies $1,2$:
\begin{equation}
    \delta\hat\varphi_{-2}=-\frac{5}{168}(d_1-d_2)^2\,.
\end{equation}
In sGB gravity, neutron stars have vanishing scalar charge, while for a BH $d\propto\zeta=\alpha/M^2$; in the case of shift-symmetric sGB gravity $f=\phi$, $d=\zeta$.

Therefore, the most significant observable quantity characterizing BHs in sGB (and sGBR) gravity is the dimensionless scalar charge $d$~\eqref{eq:dsc}. Since in theories not allowing scalarization $d\propto\zeta$, and thus the observable effect increases for smaller BH masses, one may naively expect that the same occurs  in theories with scalarization, and then that the couplings allowing for smaller BH masses have a potentially more interesting phenomenology. However, it should be noted that --~in theories allowing scalarization, like those considered in this article~-- the dependence of the dimensionless scalar charge on the dimensionless coupling, $d(\zeta)$, is more complicated. Indeed $d$ depends on the couplings of the nonlinear interactions that quench the tachyonic instability associate to scalarization, while $\zeta$ controls the onset of that instability~\cite{Macedo:2019sem,Antoniou:2021zoy}.

We have computed the dimensionless scalar charge in sGB and sGBR gravity with Gaussian coupling~\eqref{eq:exponentialcoupling}, for different values of $\gamma$ an $\beta$. In Fig.~\ref{fig:QMvsM_alpha0} we show
$d$ as a function of $\hat M$ for different values of $\gamma$. In Fig.~\ref{fig:QMvsM} we show the same function for two values of $\gamma$ (corresponding to the left and right panels), and for different values of $\beta$. We can see that for each value of the parameters, the dimensionless scalar charge is bounded by a maximum value, $d\le d_{\rm max}$. 

Although $d_{\rm max}$ increases as $\gamma$ increases, it never reaches values larger than $2.5$ in sGB gravity. The Ricci coupling does not significantly affect this picture. This can be clearly understood  from Fig.~\ref{fig:QMvsGamma}, where $d_{\rm max}$ and the charge $d$ for $M=M_{\rm thr}$ are shown as functions of $\gamma$, for sGB gravity (left panel) and for sGBR gravity with $\beta=1$ (right panel). We can see that $d_{\rm max}$ is always smaller than $\simeq2.5$ in sGB gravity. In sGRB gravity, larger values can be obtained, but only in the unphysical elliptic region (see also Fig.~\ref{fig:QMvsM}).



\section{Discussion and conclusions}\label{sec:concl}
In this article we considered the hyperbolicity of the perturbation equations for static BH solutions in sGB gravity, focussing on the class of generalized Gaussian couplings~\eqref{eq:exponentialcoupling}
\begin{equation}
\label{eq:exponentialcoupling1}
    f(\phi) = \frac{1}{8 \gamma}\left(1-e^{
    - \gamma \phi^2}\right)\,,
\end{equation}
which are useful to probe the theoretical and phenomenological limitations of sGB gravity. We also consider the effect of the coupling of the scalar field with the Ricci scalar (sGBR gravity), with $J(\phi)=\phi^2$.
We find that the minimum mass of BHs under which the perturbations of static solutions become elliptic, can be chosen to be arbitrarily small in this class of couplings, since $\hat{M}_{\rm thr}\propto \gamma^{-1}$.

A possible explanation of this behavior may rely on the fact that in the coupling functions~\eqref{eq:exponentialcoupling}, the dimensionful coupling constant $\alpha$ is divided by \bb. Thus
the cut-off scale of the EFT is  proportional to $\sqrt{\gamma}$.
This does not explain why the threshold mass is proportional to $\gamma^{-1}$, but may provide an indication of such behaviour.

Although larger values of $\gamma$ allow for smaller BHs, which have larger curvature, they do not correspond to larger dimensionless scalar charge. Increasing $\gamma$, then, does not lead to larger effects on the gravitational waveforms of BH binary coalescences.

In the case of sGBR gravity, we extend the analysis of~\cite{Antoniou:2021zoy,Antoniou:2022agj}, where it was noted that for some choices of the coupling parameters, the Ricci coupling improves the hyperbolicity properties of the perturbation equations. We find that in the case of  generalized Gaussian coupling, this conclusion still holds for small values of the coupling parameter $\beta$, for which the threshold mass is smaller than for $\beta=0$.
For larger values of this parameter we find instead the opposite effect: the Ricci coupling makes the threshold mass larger than in the corresponding sGB theory. We note that this result is not in tension with the the results of Ref.\,\cite{Thaalba:2023fmq}, as the setup is different --- collapse rather than black hole perturbations and different coupling functions to the GB term. In combination, our analysis and that of Ref.\,\cite{Thaalba:2023fmq} indicate that the effect of the Ricci coupling on the hyperbolicity of the theory is complex, and depends on the specific coupling considered and the setup.

\acknowledgments
We thank G. Antoniou, C. Macedo for their help in comparing our results, and M. Della Rocca for useful comments and discussions.
D.R. and L.G. acknowledge financial support from the EU Horizon2020 Research and Innovation Programme under the Marie Sklodowska-Curie Grant Agreement No. 101007855. T.P.S. acknowledges partial support from the STFC Consolidated Grant nos. ST/V005596/1,  ST/X000672/1, and UKRI2492. 

\appendix

\section{\bbcrit~analysis}

\label{appendix:beta_crit}
As discussed in Sec.~\ref{subsec:background}, stationary black holes in sGB gravity with generalized Gaussian coupling~\eqref{eq:exponentialcoupling} can exist for arbitrary small masses as long as the parameter $\gamma$ is larger than a critical value $\gamma_{\rm crit}$. We here discuss how this value is affected by the introduction of the Ricci coupling.

\subsubsection*{sGB gravity}

Let us first review the case of sGB gravity. A necessary condition for the existence of a static black hole solution is the regularity of the scalar field and of its derivative at the horizon. This in only possible as long as the regularity condition~\cite{Blazquez-Salcedo:2016enn,Kanti:1997br,Kanti:1997br,Doneva:2017bvd,Silva:2017uqg}
\begin{equation}
\label{eq:f_regularity}
    \freg(\phi_h)= r_h^4 - 96\alpha^2 (f'(\phi_h))^2>0,
\end{equation}
where $\phi_h$ is the value of the scalar field at the horizon $r=r_h$.

\begin{figure}[t]
    \centering
        \includegraphics[width=0.8\linewidth]{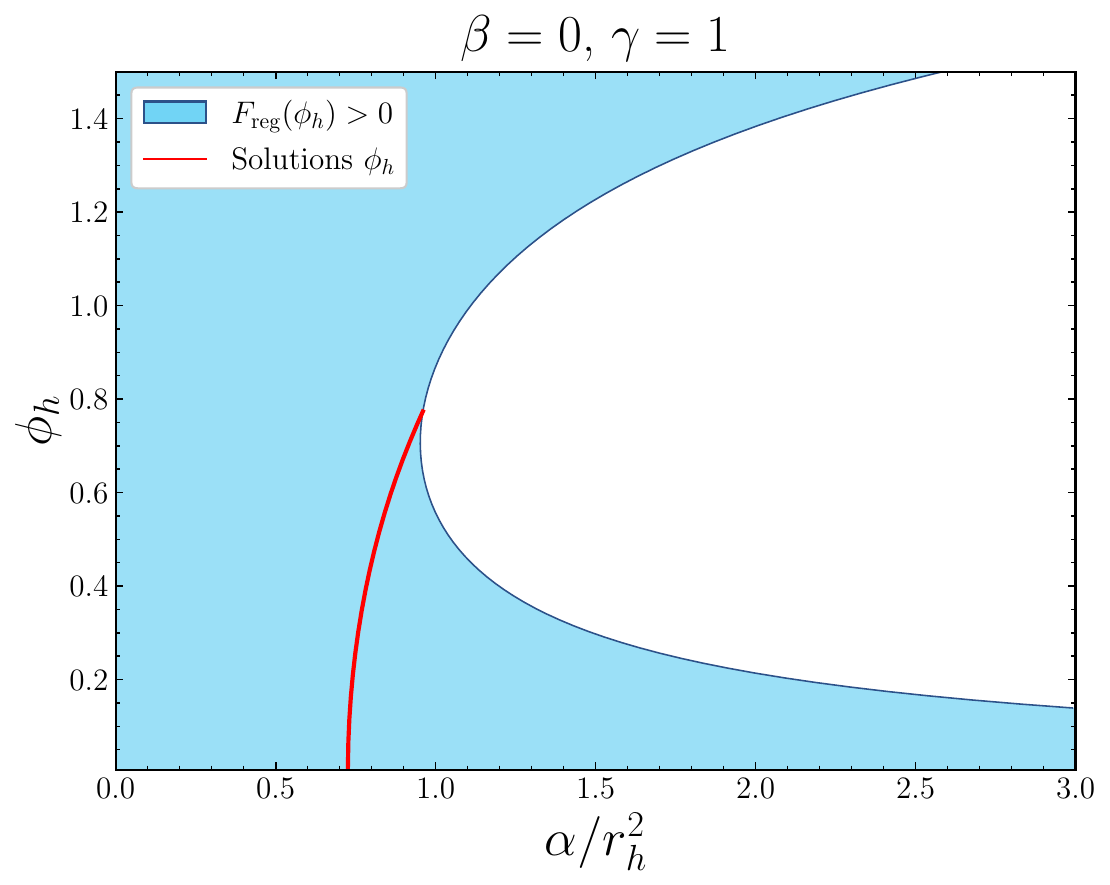}\\
        \label{fig:img_a}
        \includegraphics[width=0.8\linewidth]{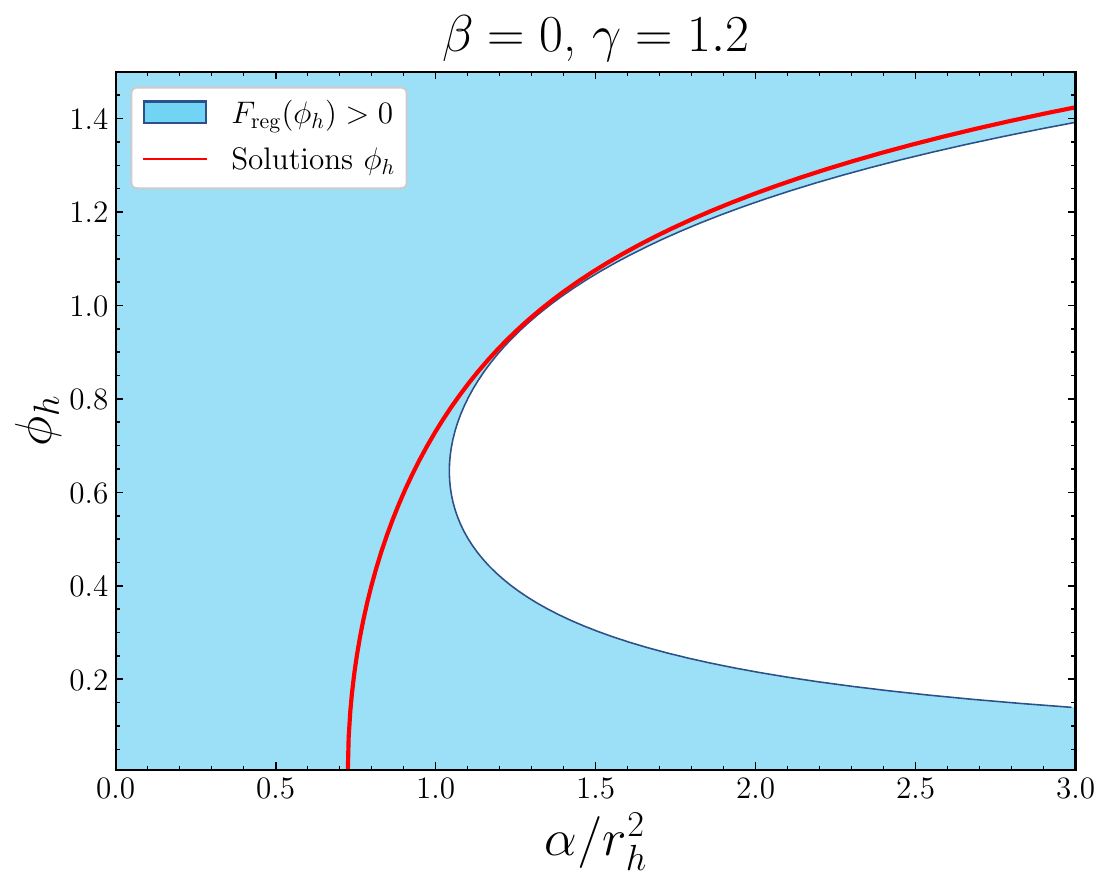}
        \label{fig:img_b}
    \caption{The shadowed regions in the $\alpha/r_h^2 - \phi_h$ plane correspond to allowed scalarized static black hole configurations in sGB gravity for which the regularity condition~\eqref{eq:f_regularity} is satisfied. The red curves correspond to configurations with vanishing scalar field at infinity, which are those we consider as physically acceptable solutions. In the top panel $\gamma<\gamma_{\rm crit}$, and $\alpha/{r^2_h}$ is bounded from above since the red curve ends in the forbidden region. In the bottom panel, instead, $\gamma>\gamma_{\rm crit}$, and $\alpha/{r^2_h}$ can reach arbitrary large values, because the red curve overcomes the forbidden region.}
    \label{fig:f_regs_positivityregions_comparison}
\end{figure}

In Fig.~\ref{fig:f_regs_positivityregions_comparison} we show the region in the the $\alpha/r_h^2 - \phi_h$ plane for which the regularity condition\,\eqref{eq:f_regularity} is satisfied (shadowed region). The solutions satifying the asymptotic condition of vanishing scalar field are represented by the red curve. We can see that if $\gamma<\gamma_{\rm crit}$ (top panel), the red curve ends in the forbidden region, leading to a maximum value of $\alpha/r_{\rm h}^2$. If, instead, $\gamma>\gamma_{\rm crit}$ (bottom panel), the curve avoids the forbidden region, and $\alpha/r_{\rm h}^2$ can reach arbitrarily large values, i.e., for a given coupling $\alpha$, the scalarized black hole can be arbitrarily small.


Thus, there is a minimum value of the parameter $\gamma$, that we call $\gamma_{\rm crit}$, for which a complete scalarized branch is allowed~\cite{Fernandes:2022kvg}. For sGB gravity with generalized Gaussian coupling we find\,\footnote{Our result has a discrepancy $\lesssim2\%$ with that of~\cite{Fernandes:2022kvg}, which we think is due to differences in the numerical implementation.}
\begin{equation}
    \label{def:beta_crit_sGB}
    \gamma_{crit} = 1.18677\,.
\end{equation}
\subsubsection*{sGBR gravity}
In sGBR gravity, the regularity condition becomes
\begin{equation}
\label{def:f_reg_2}
    \begin{split}
        \freg(\phi_h) &= (J(\phi_h)+1)^2 \Big(-1152 \alpha ^3 f'(\phi_h)^3 J'(\phi_h)- \\ & \quad  -48 \alpha ^2 f'(\phi_h)^2 \left(9 J'(\phi_h)^2+2\right)+\\ & \quad +J(\phi_h) \left(-96 \alpha ^2 f'(\phi_h)^2+6 J'(\phi_h)^2+2\right)+ \\ & \quad +\left(3 J'(\phi_h)^2+1\right)^2+J(\phi_h)^2\Big)>0\,.
    \end{split}
\end{equation}
We note that the regularity function $\freg$ identically vanishes for $J(\phi_{h})=-1$. In the case of  the quadratic coupling function $J=-\beta\phi^2/4$ (Eq.~\eqref{def:coupling_Ricci}), this occurs for
\begin{equation}
\label{eq:fixed zero}
    \phi_h = 2\sqrt{\frac{1}{\beta}}\,,
\end{equation}
regardless of the value of $\gamma$. This explains while even for large values of $\gamma$, scalarized black hole solutions do not exist for arbitrarily small values of the mass, as shown in Fig.~\ref{fig:QvsM_stability}.


This motivates us to consider a different coupling to the Ricci scalar such that $J(\phi_{h})\neq-1$, for all values of $\phi_h$. This is the case of the Gaussian coupling of Eq.~\eqref{eq:Ricci_exp_coupling}, analyzed in Appendix~\ref{appendix:ricciexpcop}.




\section{Gaussian Ricci coupling}
\label{appendix:ricciexpcop}
We here consider a different choice of the Ricci coupling function: the Gaussian coupling
\begin{equation}
\label{eq:Ricci_exp_coupling}
    J(\phi)=- \frac{\beta}{4}\left(1- e^{-\phi^2}\right).
\end{equation}
In this case for $\beta<4$, $J(\phi_{h})\neq-1$ for all values of $\phi_{\rm h}$, and thus --~as we are going to show~-- scalarized configurations exist for arbitrarily small masses, as long as $\gamma$ is sufficiently large.

\begin{figure}[t]
    \centering
    \includegraphics[width=0.8\linewidth]{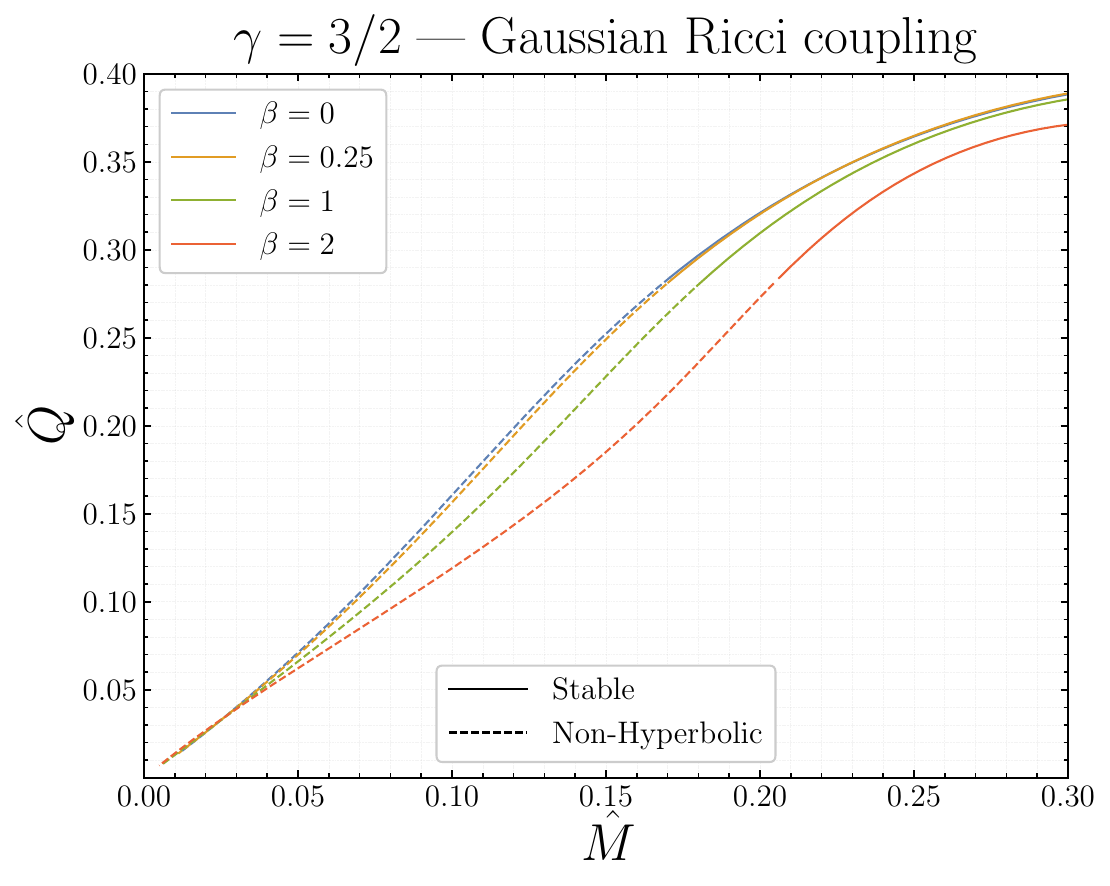}
    \includegraphics[width=0.8\linewidth]{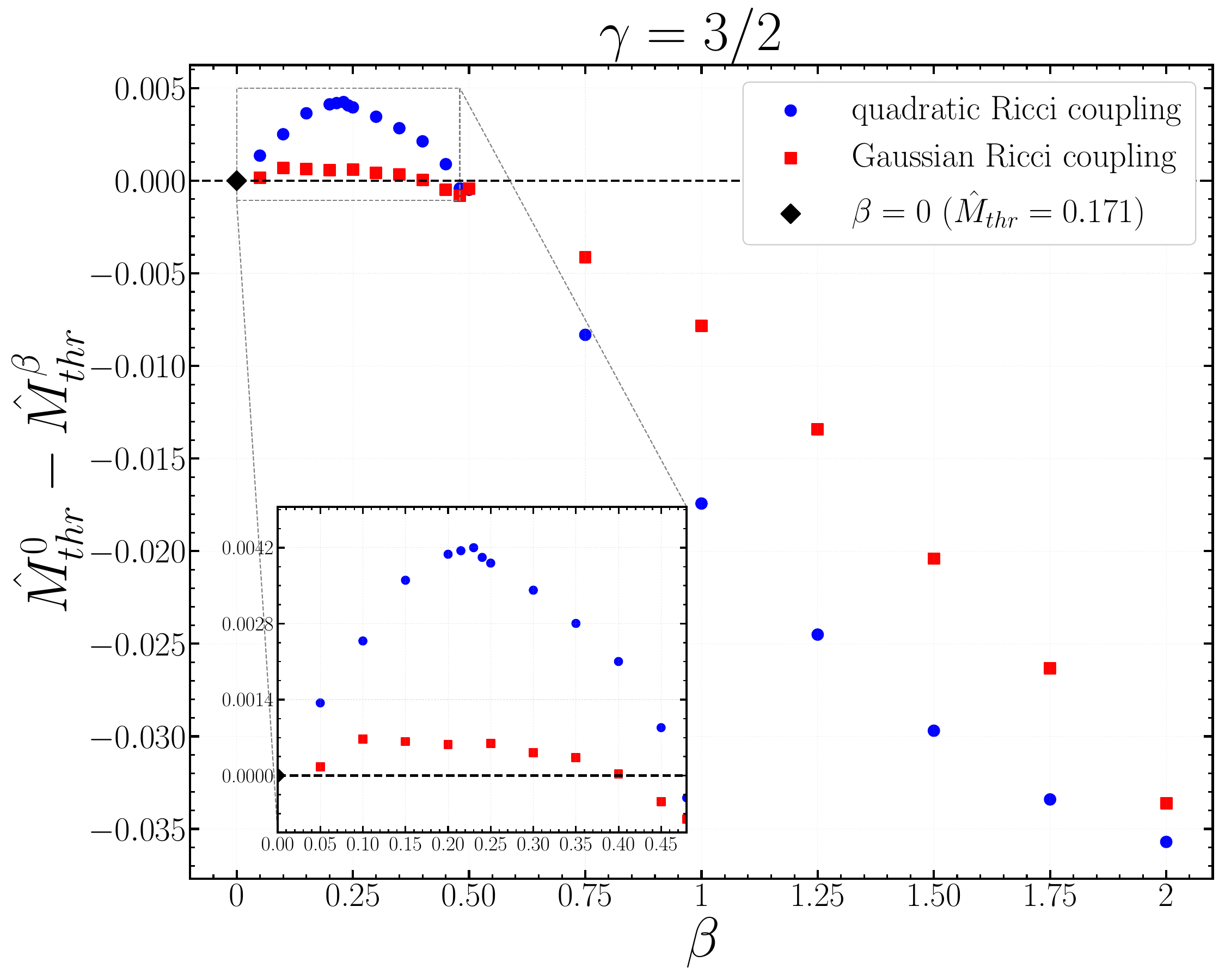}
    \caption{Same as Fig.~\ref{fig:QvsM_stability} but with Gaussian coupling~\eqref{eq:Ricci_exp_coupling} to the Ricci scalar.  We can see that, as in the case of quadratic Ricci coupling, the Ricci coupling increases (decreases) the threshold mass for $\beta$ larger (smaller) than $\simeq0.50$.}  
    \label{fig:QvsM_stability_expRicci}
\end{figure}
\begin{figure}[t]
    \centering
    \includegraphics[width=0.8\linewidth]{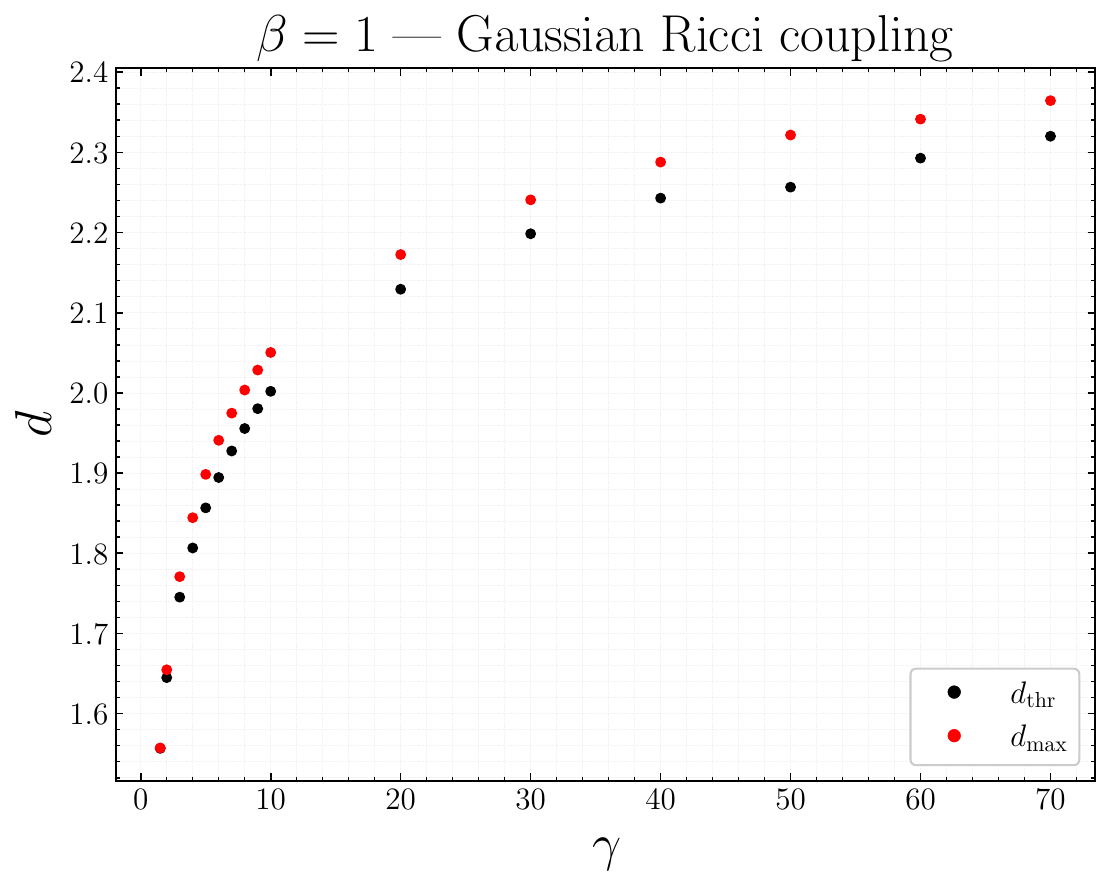}
    \caption{Same as the right panel of Fig.~\ref{fig:QMvsGamma}, but with Gaussian Ricci coupling Eq.~\eqref{eq:Ricci_exp_coupling}.}
    \label{fig:QMvsGamma_expRicci}
\end{figure}
\begin{table}[t]
  \centering
  \begin{tabular}{c|c|c}
    \hline
    $\beta$ & $b$ & $c$ \\
    \hline
    0.0 & 0.2337 & -0.00012 \\
    0.25 & 0.2379 & -0.000069 \\
    0.40 & 0.2411 & -0.000078 \\
    0.50 & 0.2437 & -0.000085 \\
    0.75 & 0.2516 & -0.000129 \\
    1.00 & 0.2612 & -0.000194 \\
    1.25 & 0.2733 & -0.000288 \\
    1.50 & 0.2865 & -0.000394 \\
    1.75 & 0.3027 & -0.000536 \\
    2.00 & 0.3179 & -0.00073 \\
    \hline
  \end{tabular}
  \caption{Values of the best fit parameters of \hMthr as a function of $1/\gamma$ for different values of $\beta$ with Gaussian coupling~\eqref{eq:Ricci_exp_coupling} of the scalar field to the Ricci scalar. The fitting model function is $ \hat{M_{\rm thr}}(\gamma)= b \gamma^{-1} + c$.}
    \label{tab:fit_params_exp}
\end{table}
\begin{figure}[t]
    \centering
    \includegraphics[width=0.8\linewidth]{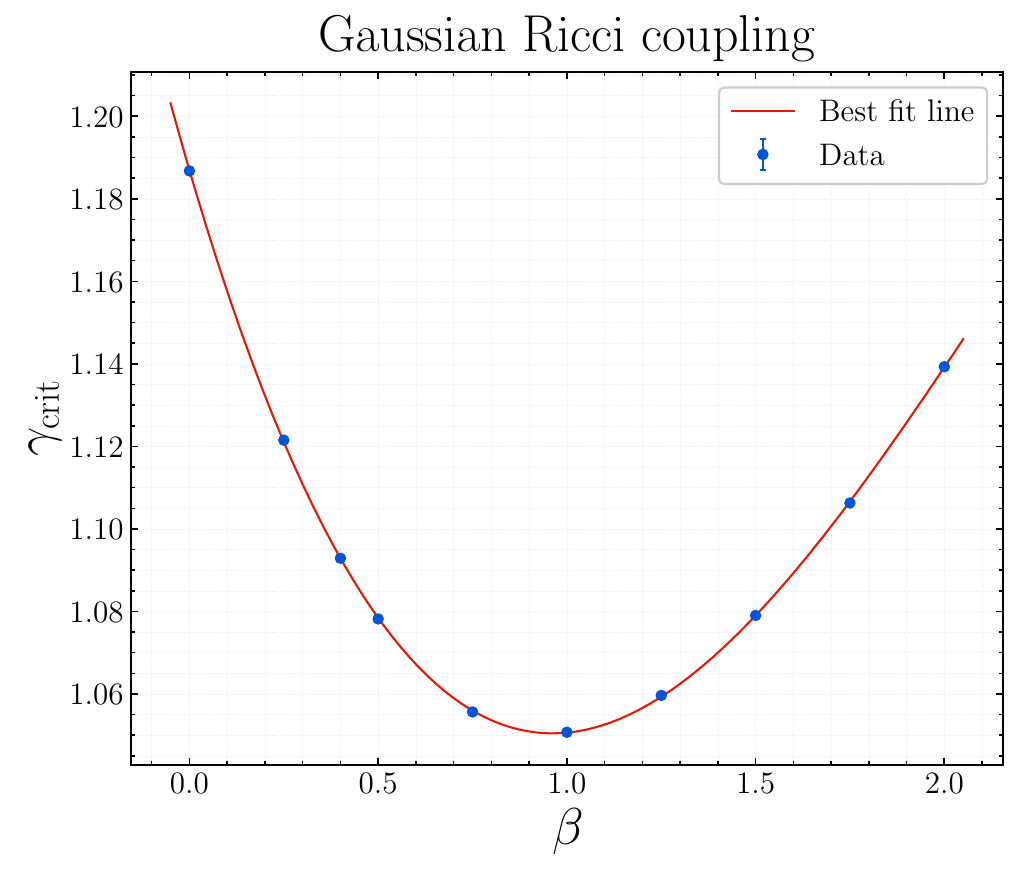}
    \caption{Dependence of $\gamma_{crit}$ on $\beta$ for Gaussian Ricci coupling of Eq.~\eqref{eq:Ricci_exp_coupling}. The red line is the
    best fit curve with model $a x^3 + b x^2 + c x +d$ and whose best parameters found are $a=-0.03258714$, $b=0.21017729$, $c= -0.31388424 $, $d= 1.18691072$.}
    \label{fig:betacritVSalpha_exp}
\end{figure}

In Fig.~\ref{fig:QvsM_stability_expRicci} we show, in the top panel, the dimensionless scalar charge $\hat{Q}$ as a function of $\hM$, for different values of $\beta$ and for $\gamma=3/2$ (larger than the critical value). The solid (dashed) curves correspond to hyperbolic (elliptic) configurations. In the bottom panel we show the shift in the threshold mass due to the Ricci coupling, as a function of $\beta$. A comparison with Fig.~\ref{fig:QvsM_stability}, which corresponds to the quadratic Ricci coupling, show the same qualitative behaviour. The main difference is that for the Gaussian Ricci coupling, stationary scalarized configurations exist for arbitrarily small values of the mass.

%

We have performed the same analysis of $\hat{M}_{\rm thr}$ as a function of $1/\gamma$ previously done for the quadratic Ricci coupling (see Fig.~\ref{fig:multiple_fits} and Tab.~\ref{tab:fit_params}). We find that the behavior is still approximately linear, and best fit parameters are displayed in Tab.\ref{tab:fit_params_exp}.

The behavior of the dimensionless scalar charge $d$ as a function of $\gamma$ for the  Gaussian Ricci coupling is shown in Fig.~\ref{fig:QMvsGamma_expRicci} for fixed $\beta=1$. As in the case of quadratic Ricci coupling, the dimensionless charge reaches an asymptotic value as $\gamma$ increases.

Finally, we study how $\gamma_{\rm crit}$ depends on the coupling constant $\beta$ in the case of Gaussian Ricci coupling. In Fig.~\ref{fig:betacritVSalpha_exp} we see that the presence of the Ricci coupling  reduces the value of $\gamma_{\rm crit}$. This effect is most significant for $\beta\simeq1$, while for larger values of $\beta$ the value of $\gamma_{\rm crit}$ increases. The function $\gamma_{\rm crit}(\beta)$ is well described by a cubic fit, whose  best parameters are provided in the caption of Fig.~\ref{fig:betacritVSalpha_exp}. 

\bibliography{biblio}
\end{document}